\documentclass{article}
\usepackage{amsmath,amsfonts,amssymb,amsthm,epsfig,epsf,psfrag}
\usepackage[nolists]{endfloat}
\usepackage{natbib}  % Kathy added
\usepackage{subfig}
\usepackage{color}
\usepackage{bm}

\definecolor{olive}{rgb}{0.3, 0.4, .1}
\definecolor{fore}{RGB}{249,242,215}
\definecolor{back}{RGB}{51,51,51}
\definecolor{title}{RGB}{255,0,90}
\definecolor{dgreen}{rgb}{0.,0.6,0.}
\definecolor{gold}{rgb}{1.,0.84,0.}
\definecolor{JungleGreen}{cmyk}{0.99,0,0.52,0}
\definecolor{BlueGreen}{cmyk}{0.85,0,0.33,0}
\definecolor{RawSienna}{cmyk}{0,0.72,1,0.45}
\definecolor{Magenta}{cmyk}{0,1,0,0}
\definecolor{britishracinggreen}{rgb}{0.0, 0.26, 0.15}
\definecolor{bulgarianrose}{rgb}{0.28, 0.02, 0.03}
\definecolor{burntorange}{rgb}{0.8, 0.33, 0.0}
\definecolor{cadmiumgreen}{rgb}{0.0, 0.42, 0.24}
\definecolor{capri}{rgb}{0.0, 0.75, 1.0}
\definecolor{ceruleanblue}{rgb}{0.16, 0.32, 0.75}
\definecolor{darkblue}{rgb}{0.0, 0.0, 0.55}
\definecolor{darkslateblue}{rgb}{0.28, 0.24, 0.55}
\definecolor{royalpurple}{rgb}{0.47, 0.32, 0.66}
\definecolor{violet(ryb)}{rgb}{0.53, 0.0, 0.69}
\definecolor{normaltextcolor}{RGB}{33,26,82}% dark blue

\def\bSig\mathbf{\Sigma}
\def\s2{\sigma^2}
\def\su2{\sigma_u^2}

\def\V{\textrm{Var}}
\newcommand{\tr}{\mbox{tr}}
\def\mb{\mathbf}

\def\bs{\boldsymbol}
\def\j{\bs{j}}
\def\T{\textsf{T}}

\def\ith{$i^{\textrm{th }} $}
\def\DaOpt{$D_{\textrm{A}}$--optimal}
\def\COpt{$C$--optimal}
\def\diag{\textrm{diag}}

\newcommand*{\bigCdot}{\raisebox{-0.25ex}{\scalebox{1.2}{$\cdot$}}}

\newcounter{examplecounter}
\newenvironment{example}{\begin{quote}%
		\refstepcounter{examplecounter}%
		\textbf{Example \arabic{examplecounter}}%
		\quad
	}{%
\end{quote}%
}

\title{Optimal block designs for experiments with responses drawn from a Poisson distribution}

\author
{Stephen Bush (Stephen.Bush@uts.edu.au) \\
Department of Mathematical and Physical Sciences, \\
University of Technology Sydney, Australia 
\and
Katya Ruggiero (k.ruggiero@auckland.ac.nz) \\
Department of Statistics, \\
University of Auckland, New Zealand}

\begin{document}

\maketitle

\begin{abstract}
Optimal block designs for additive models achieve their efficiency by dividing experimental units among relatively homogenous blocks and allocating treatments equally to blocks. Responses in many modern experiments, however, are drawn from distributions such as the one- and two-parameter exponential families, e.g., RNA sequence counts from a negative binomial distribution. These violate additivity. Yet, designs generated by assuming additivity continue to be used, because better approaches are not available, and because the issues are not widely recognised. We solve this problem for single-factor experiments in which treatments, taking categorical values only, are arranged in blocks and responses drawn from a Poisson distribution. We derive expressions for two objective functions, based on $D_\textrm{A}$- and $C$-optimality, with efficient estimation of linear contrasts of the fixed effects parameters in a Poisson generalised linear mixed model (GLMM) being the objective. These objective functions are shown to be computational efficient, requiring no matrix inversion. Using simulated annealing to generate Poisson GLMM-based locally optimal designs, we show that the replication numbers of treatments in these designs are inversely proportional to the relative magnitudes of the treatments' expected counts. Importantly, for non-negligible treatment effect sizes, Poisson GLMM-based optimal designs may be substantially more efficient than their classically optimal counterparts.

\end{abstract}

%  Please place your key words in alphabetical order, separated
%  by semicolons, with the first letter of the first word capitalized,
%  and a period at the end of the list.
%

%  As usual, the \maketitle command creates the title and author/affiliations
%  display 

%  If you are using the referee option, a new page, numbered page 1, will
%  start after the summary and keywords.  The page numbers thus count the
%  number of pages of your manuscript in the preferred submission style.
%  Remember, ``Normally, regular papers exceeding 25 pages and Reader Reaction 
%  papers exceeding 12 pages in (the preferred style) will be returned to 
%  the authors without review. The page limit includes acknowledgements, 
%  references, and appendices, but not tables and figures. The page count does 
%  not include the title page and abstract. A maximum of six (6) tables or 
%  figures combined is often required.''

%  You may now place the substance of your manuscript here.  Please use
%  the \section, \subsection, etc commands as described in the user guide.
%  Please use \label and \ref commands to cross-reference sections, equations,
%  tables, figures, etc.
%
%  Please DO NOT attempt to reformat the style of equation numbering!
%  For that matter, please do not attempt to redefine anything!

\section{Introduction}
\label{s:intro}

The introduction of gene expression microarrays \citep{SSDB95} towards the end of the twentieth century initiated the start of a biotechnology revolution of rapidly evolving instruments capable of profiling a wide range of different molecular species at the cellular level. Today, high resolution instruments, such as next-generation sequencing (NGS) technologies \citep{craig2008identification}, are capable of generating counts of individual copies of, for example, different gene transcripts. Experiments using these technologies are relatively expensive, resulting in studies with low numbers of biological replicates of treatments, making the efficient statistical design of such experiment critical. Yet, with few exceptions, classically optimal designs based on the assumption of unit-treatment additivity (i.e. functional independence of variances and covariances on their means) continue to be used. 

Optimal statistical designs are central to conducting efficient comparative experiments, enabling contrasts of treatment parameters to be estimated without bias and minimum variance while requiring minimum effort, subjects, or other resources. A rich body of literature on optimal designs \citep{JW1995,ADT2007} has grown since the creation of this field of statistics \citep{KS1918}. Until very recently however, the criteria used to define optimal designs has depended on the assumption of unit-treatment additivity. 

Under additivity, optimal designs achieve their efficiency by dividing the experimental material into relatively homogeneous blocks and allocating treatments equally to blocks. Full efficiency, whereby a model's treatment parameters can be estimated independently of its block parameters, is attained when treatments can be arranged in a complete block design \citep{F1926}, i.e. each treatment occurs equally frequently, usually once, in each block. However, as already noted, responses in
many modern experiments are drawn from distributions such as the one and two parameter exponential families, e.g., RNA sequence counts from a negative binomial distribution \citep{AD2010}. These violate additivity. Our current work on single-factor experiments shows that optimal block designs from the classical setting can be importantly non-optimal when additivity is violated.

\citet{C1988} was arguably the first to consider the problem of randomised experiments in the context of responses drawn from an exponential family distribution, with the objective being the difference between treatment groups in the canonical parameter. The approach taken by Cox is to use conditioning to eliminate the blocking effects, and argued that local arguments can be made when in the presence of small effects and when asymptotic maximum likelihood theory is reasonable. The author also suggests that treating blocking as a random variable, which is the approach of this paper, is reasonable.

The last decade has witnessed increasing interest in research on methods for the optimal design of experiments based on the structure of general exponential distributions.  In \citeyear{khuri2006design}, \citeauthor{khuri2006design} presented a comprehensive review of design issues for generalised linear models (GLMs) in the absence of any random parameters, discussing the dependence of optimal designs on the values of the canonical parameters of the model; a problem which persists for generalised linear mixed models. Using compromise criteria, \citet{woods2006designs} proposed a method for finding exact designs robust to misspecification of the model's functional form for experiments involving several explanatory variables, where the factors take values along a bounded continuum. Exact designs constrain the weights placed on each treatment combination so that, for a given sample size, the replication number of each treatment (combination) is an integer (following the terminology of \citet{ADT2007}). In contrast, continuous designs do not use this constraint, and replication numbers are obtained by making nearest integer approximations. \citet{russell2009d} present results for generating D-optimal continuous designs for Poisson regression where there are several continuous bounded factors.  They also discuss the implementation of compromise designs to obtain designs that are robust to parameter misspecification. \citet{niaparast2013optimal} investigated Poisson regression with a continuous predictor and random intercept. They argued that finding an optimal design using standard likelihood methods is cumbersome, even for simple models. 

The optimal block designs of experiments with multiple bounded continuous factors and correlated non-normal responses was first considered by \citet{woods2011blocked}. They used generalised estimating equations to incorporate block effects into the variance estimate for fixed effect parameters for GLMs. They considered both exchangeable and autoregressive correlation patterns, and presented two strategies for constructing block designs. The first strategy uses simulated annealing (\citet{kirkpatrick1984optimization}, \citet{haines1987application} and \citet{woods2010robust}), while the second allocates the runs of the optimal unstructured design to blocks in an optimal way.

\citet{yang2015optimal} present results that give D-optimal continuous factorial designs for logistic regression, and consider the use of exchange algorithms to find D-optimal exact designs in the absence of blocking variables.

To date, work in developing methods for generating optimal block designs for experiments with responses drawn from an exponential family distribution has been carried out exclusively in the context of response surface models. We consider, in contrast, block designs for the much more widely applicable class of designs in which the values of the factor levels are fixed at the outset and play no role in design optimality. As far as we are aware, there are currently no methods available for generating optimal block designs in this setting. Yet, as demonstrated by next-generation sequencing experiments, there is a very real and pressing need for such methods. 

In this paper, we develop methods for generating optimal block designs for category-valued single factor-experiments with responses drawn from a Poisson distribution, and with efficient estimation of contrasts of the model parameters being the objective. In Section~\ref{sec:models} we develop the notation and definitions needed to specify the generalised linear mixed model (GLMM) for responses drawn from an exponential family of distributions and the corresponding pseudo-likelihood estimating equations. From these we derive the marginal Fisher information matrix for the estimation of the fixed effects in the model. In Section~\ref{sec:GLMMs-CorrCount}, we develop objective functions based on $D_\textrm{A}$-- and \COpt{ity} for the efficient estimation of the fixed effects parameters in a Poisson GLMM. While these optimality criteria generally result in objective functions which require matrix inversion, we show that for Poisson GLMMs with log link the objective functions can be simplified so that less inversion is necessary, leading to computational efficiency in the search for optimal designs. We use simulated annealing to search the set of competing designs for experiments of a given size, with the search space constrained to those designs which are locally optimal based on point prior estimates of the fixed effect parameters. Our key inputs into the simulated annealing algorithm are described in Section~\ref{sec:simulated-annealing}. In Section~\ref{sec:example} we consider two examples, including a next-generation sequencing experiment, where we generate locally optimal block designs using the methods developed in Sections~\ref{sec:GLMMs-CorrCount} and~\ref{sec:simulated-annealing}. These show that, for a fixed number of blocks with constant block size, the replication of treatments in Poisson GLMM-based optimal designs are inversely proportional to the relative magnitudes of the treatments' expected counts, which flies in the face of our traditional belief of optimality being achieved through the (near-) balanced allocation of treatments across blocks. They further show that, for experiments with non-negligible effect sizes, the Poisson GLMM-based optimal designs may be substantially more efficient than optimal designs from the classical setting assuming additivity.

With these methods in hand, experimenters will be enabled in correctly answering their research questions with minimum effort, subjects, or other resources.

\section{Models}
\label{sec:models}

In this section we introduce generalised linear mixed models as an extension of both generalised linear models and of linear mixed models. As we progress to derive expressions for design optimality criteria, we observe that features of efficient designs for both generalised linear models and of linear mixed models are present in efficient designs for generalised linear mixed models. 

\subsection{The generalised linear model}
\label{subsec:GLMs}

Consider an experiment in which $t$ treatments are arranged in a completely randomised design comprising $n$ experimental units, where $n$ is a multiple of $t$.  We define a linear model for an $n \times 1$ vector of observations as
\begin{align}
\label{eq:LM}
\bs{y} &= X \bs{\beta} + \bs{e},
\end{align}
where $\bs{\beta} = (\alpha, \bs{\tau}^{\T})^{\T}$ is a $p \times 1$ vector of parameters containing the fixed effects $\alpha$, denoting the overall mean of the observations, and $\bs{\tau}=(\tau_1,\ldots,\tau_t)^\T$, denoting the $t$ treatment effect parameters. The $n \times p$ treatment design matrix, $X$, characterises the allocation of treatments to experimental units and, therefore, the fixed effect parameters associated with each observation in $\bs{y}$. The $n\times 1$ vector of residual errors, $\bs{e}$, is assumed to be independently and identically distributed normal with constant variance. If the response variable does not give rise to this error distribution, then an alternative model needs to be considered. One such alternative is the generalised linear model (GLM). 

GLMs are used when the responses in $\bs{y}$ are assumed to arise from a distribution belonging to the exponential family of distributions. Such distributions include, for example, the binomial distribution for binary responses and the Poisson distribution for count responses. In a GLM the vector of mean responses, $\bs{\mu}$, and the linear predictors, $X\bs{\beta}$, are related by a canonical link function, $g(\bigCdot)=b'(\bigCdot)^{-1}$. This gives rise to the model form  
\begin{eqnarray}
\label{eq:link-fn}
\bs{\eta} = g(\bs{\mu}) = X\bs{\beta}.
\end{eqnarray}
It follows that $E(\bs{y}) = \bs{\mu} = b'(\bs{\theta})$ and $\V(\bs{y}) = b''(\bs{\theta})a(\phi)$, where $b(\bs{\theta})$ and $a(\phi)$ denote functions of the natural parameter, $\bs{\theta}$, and the dispersion parameter, $\phi$, respectively, for the independent observations in $\bs{y}$. 

\subsubsection{Fisher information matrix for GLMs}
\label{subsub:GLM-EstEqns}

The maximum likelihood estimator, $\widehat{\bs{\beta}}$, of the fixed model effects, $\bs{\beta}$, in a GLM is asymptotically normally distributed. The covariance matrix of $\widehat{\bs{\beta}}$ is the inverse of the Fisher information matrix which is derived from the log-likelihood function of the GLM, i.e.
\begin{equation*}
\label{eq:glm-infoMat}
M(\xi,\bs{\beta}) = E\Bigg\{-\frac{\partial^2\ell(\bs{\theta})}{\partial\bs{\beta}^2} \Bigg\} = E\Bigg\{-\frac{\partial^2\ell\Big[\theta\big(g^{-1}(X\bs{\beta}; y, \phi)\big)\Big]}{\partial\bs{\beta}^2} \Bigg\} = X^\T WX,
\end{equation*}
where $W = (DVD)^{-1}$, $V=\diag[\textrm{Var}(y_i)]$ and $D=\diag[\partial\eta_i/\partial\mu_i]$. Hence, the information matrix depends on the link function, since $\partial\eta_i/\partial\mu_i=g'(\mu_i)$,  the design, $\xi$, through the design matrix $X$ and the parameters in $\bs{\beta}$.

\subsubsection{Poisson GLM}
\label{subsub:GLMMs-PoissonModel}

Responses in many modern experiments are drawn from distributions such as the one- and two-parameter exponential families, e.g., RNA sequence counts from a negative binomial distribution. Here we focus exclusively on experiments in which responses from the \ith treatment group are counts independently drawn from the one-parameter Poisson distribution, i.e. $\bs{y}_i\sim \textrm{Poisson}(\bs{\lambda}_i)$, with canonical link function $g(\bigCdot)=\log(\bigCdot)$.

\subsection{The linear mixed model}
\label{subsec:LMMs}

Consider now an experiment in which $t$ treatments are arranged in a generalised block design with $b$ blocking factors. We define the linear mixed model (LMM) for an $n\times 1$ vector of observations, $\bs{y}$, using the general matrix notation
\begin{equation}
\label{eq:lmm}
  \bs{y} = X\bs{\beta} + Z\mb u + \bs{e},
\end{equation}
where the linear component, $\bs{\mu}= X\bs{\beta}$, represents the expected responses of the marginal model, with fixed effects parameter vector, $\bs{\beta}$, and treatment design matrix, $X$, defined as in (\ref{eq:LM}).  The vector of block random effect parameters $\bs{u}=(\bs{u}_1^\T,\ldots, \bs{u}_b^\T)^\T$ is multivariate normally (MVN) distributed, with sub-vector $\bs{u}_i=(u_{i1},\ldots, u_{ib_i})^\T\sim MVN(\bs{0},G_i)$, where $G_i=\s2_iI$, corresponding to the $i$th block factor, $i=1,\ldots,b$. The $n\times b$ %_{\bigCdot}$ 
block design matrix, $Z$, characterises the association of experimental units and, therefore, random effect parameters with each
observation in $\bs{y}$. Finally, the $n\times 1$ vector of residual error parameters $\bs{e}\sim \textrm{MVN}(0, R)$. In the following, we consider only the case where these errors are uncorrelated, i.e. $R=\s2I$.

\subsubsection{Fisher information matrix for LMMs}
\label{subsub:LMM-EstEqns}

The LMM estimating (or normal) equations are given by
\begin{equation}
\label{eq:LMM-EstEqns}
\begin{bmatrix}
 X^\T R^{-1} X & X^\T R^{-1} Z \\
 Z^\T R^{-1} X & Z^\T R^{-1} Z + G^{-1}
\end{bmatrix}
\begin{bmatrix}
 \bs{\beta}\\
 \bs{u}
\end{bmatrix}=
\begin{bmatrix}
 X^\T R^{-1}\bs{y} \\
 Z^\T R^{-1}\bs{y}
\end{bmatrix}.
\end{equation}

Solving the estimating equations in (\ref{eq:LMM-EstEqns}) for the fixed model effects, $\bs{\beta}$, and a design $\xi$ yields the estimate $M(\xi,\bs{\beta},\sigma,\sigma_u)\widehat{\bs{\beta}} = XV^{-1}\mathbf{y}$, where the information matrix $M(\xi,\bs{\beta},\sigma,\sigma_u)=X^\T V^{-1} X$ and the weight matrix $V=ZGZ^\T + R$.

\subsection{The generalised linear mixed model}
\label{subsec:GLMMs}

Extending either the LMM in (\ref{eq:lmm}) to allow the observed responses to arise from a distribution in the exponential family, with linear predictor defined as in (\ref{eq:link-fn}), or the GLM defined in (\ref{eq:link-fn}) to also include random effects, yields the generalised linear mixed model (GLMM)
\begin{equation}
\label{eq:y|u}
  \bs{\eta}=g(\bs{\mu})=g[\textrm{E}(\bs{y}|\bs{q})] = X\bs{\beta} + Z\mathbf{q},
\end{equation}
where $\bs{y}|\bs{q}$ denotes the vector of responses, conditional on the random effects $\bs{q}= (\bs{u}^{\T}, \bs{e}^{\T})^{\T}$, arising from an exponential family of distributions. The random effect parameter vector, $\bs{u}$, and vector of residual error parameters, $\bs{e}$, are defined as in (\ref{eq:lmm}). As expected, when the link function, $g(\bigCdot)$, is the identity, the GLMM reduces to the ordinary LMM.

\subsubsection{Fisher information matrix for GLMMs}
\label{subsub:GLMMs-EstEqs}

The pseudo-likelihood estimating equations for the GLMM defined in (\ref{eq:y|u}) are
\begin{equation}
\label{eq:pseudoLik-glmm}
\begin{bmatrix}
 X^\T W X & X^\T W Z \\
 Z^\T W X & Z^\T W Z + G^{-1}
\end{bmatrix}
\begin{bmatrix}
 \bs{\tau}\\
 \bs{u}
\end{bmatrix}=
\begin{bmatrix}
 X^\T W\bs{y^\star} \\
 Z^\T W\bs{y^\star}
\end{bmatrix},
\end{equation}\\
where $W = (DV_{\mu}^{1/2}AV_{\mu}^{1/2}D)^{-1}$ and $\bs{y^\star} = \bm\eta + (\bm y -\bm \mu) g'(\bm \mu)$ is a pseudo-variable. In general, $D = \partial \bs\mu/\partial\bs\eta$, $V_{\mu} = \diag(\sqrt{\partial^2 b(\bs \theta)/\partial \bs{\theta}^2})$ and $A = \diag(1/a(\phi))$, where $\phi$ is the scale parameter of the response distribution. The Fisher information matrix for the estimation of both the fixed and random effects is 
\begin{equation}
\label{eq:FIM-glmm}
M(\xi,\bs{\beta},\sigma,\sigma_u) = \begin{bmatrix}
X^\T W X & X^\T W Z \\
Z^\T W X & Z^\T W Z + G^{-1}
\end{bmatrix}
\end{equation}\\
\citep{stroup2012generalized}. While it may be tempting to use the conditional form of the information matrix, $X^{\T}WX$, this does not ensure that the random effects are estimable. Instead, we follow \citet{niaparast2013optimal} and \citet{WW2015} and use the marginal information matrix for the estimation of the fixed effects. 

For a generalised block design $\xi$, we partition the design matrix defined in (\ref{eq:FIM-glmm}) into four sub-matrices, i.e.
\begin{equation*}
M(\xi,\bs{\beta},\sigma,\sigma_u) = \left[\begin{array}{cc}
M_{11}(\xi,\bs{\beta},\sigma,\sigma_u) & M_{12}(\xi,\bs{\beta},\sigma,\sigma_u) \\
M_{21}(\xi,\bs{\beta},\sigma,\sigma_u) & M_{22}(\xi,\bs{\beta},\sigma,\sigma_u)
\end{array} \right] = \left[\begin{array}{cc}
M_{11} & M_{12} \\
M_{21} & M_{22}
\end{array} \right],
\end{equation*}
where sub-matrix $M_{11}$ contains the information pertaining to the fixed effects of interest, the efficiencies of which we would like to optimise, and $M_{22}$ contains the information for the remaining effects. For Poisson regression in blocks, $M_{11}$ contains the information for the fixed effects in $\bs{\beta}$ and $M_{22}$ contains the information for the random effects. Then, from results on the inverse of a partitioned matrix \citep[p. 98]{harville1997matrix}, the marginal information matrix for the estimation of $\bs{\beta}$ is given by
\begin{equation}
\label{eq:FIM-glmm-FE}
M^{\textrm{marg}}_{\bs{\beta}}(\xi,\bs{\beta},\sigma,\sigma_u) =  M_{11} - M_{12} (M_{22})^{-1} M_{21}.
\end{equation}
In the next section, we derive (\ref{eq:FIM-glmm-FE}) for Poisson regression with unstructured treatments in blocks.

%\subsubsection{Poisson GLMM for correlated count data}
%\label{subsub:GLMMs-GeneralModel}

%Following from equation (\ref{eq:y|u}), the Poisson GLMM with link function $g(\bigCdot)=\log(\bigCdot)$ mixes the Poisson log-mean with normal random effects, i.e. 
%\begin{equation*}
%\label{eq:PoissonGLMM}
%    \bs{\eta}=\log[\textrm{E}(\bs{y}|\bs{u})] = X\bs{\tau} + Z\bs{q},\tag{\ref{eq:y|u}}
%\end{equation*}
%where the random effects are, once again, defined as in (\ref{eq:lmm}). \textcolor{blue}{\bf Do we need this section? More specifically, I don't think we need this equation since it is just a repeat of (7).}

\section{Optimal block designs for correlated count data}
\label{sec:GLMMs-CorrCount}
In this section we develop objective functions for the efficient estimation of the fixed effects in a Poisson GLMM for block designs with unstructured treatments.

Consider an experiment in which $t$ treatments are arranged in $b$ blocks of equal size $k$. Assuming observations $y_{ij}$ from unit $j$ in block $i$ are conditionally Poisson-distributed with expected value given by the rate parameter $\lambda_{R(i,j)}$, where $R(i,j)\in \{1,\ldots,t\}$ denotes the label for the treatment randomised to the $(i,j)$th unit, $i=1,2,\ldots,b$ and $j=1,2,\ldots,k$. The GLMM for this situation can be written as
\begin{equation}
\label{eq:bd-glmm}
  \eta_{R(i,j)} = \alpha + \tau_{R(i,j)} + u_i + e_{ij},
\end{equation}
where $\eta_{R(i,j)}$ denotes the response on the linear predictor scale, $\alpha$ is the overall mean and $\tau_{R(i,j)}$ is the fixed effect of treatment $R(i,j)$. The block effects, 
$u_i$, are assumed to be random $N(0,\sigma_u^2)$ with $\textrm{cov}(u_i, u_{i'})=\sigma_u^2$ for $i = i'$ and zero otherwise. The residual errors, $e_{ij}$, associated with each unit are assumed $N(0,\sigma^2)$ and mutually uncorrelated.

The model specified in (\ref{eq:bd-glmm}) satisfies the GLMM definition in (\ref{eq:y|u}), where  $\bs{\eta} = [\eta_{R(i,j)}]$, $X$ is the treatment design matrix, $\bs{\tau} = (\tau_1, \cdots, \tau_t)^\T$ is a vector fixed effect treatment parameters and the vector of random effect parameters $\bs{q} = (\bs{u}^\T, \bs{e}^\T)^\T=(u_1, \cdots, u_b, e_{11}, \cdots, e_{bk})^\T$. Since all blocks are of equal size $k$, then the block design matrix $Z = (Z_b | I_n)$, where $Z_b = I_b \otimes \j_k$, $I_b$ is an identity matrix of order $b$, $\j_k$ is a $k \times 1$ vector of ones, $n = bk$, and $\otimes$ denotes the Kronecker (or outer) product.  

The Poisson GLMM with link function $g(\bigCdot)=\log(\bigCdot)$, as defined in (\ref{eq:bd-glmm}), can be expressed as a Poisson--Log-normal mixture, since $y_{ij} | \lambda_{ij} v_{ij} \sim \textrm{Poisson}(\lambda_{ij} v_{ij})$ and $\exp(v_{ij}) = e_{ij} \sim N(0, \sigma^2)$. It incorporates overdispersion through the residual parameters, $e_{ij}$, in a way that is consistent with how the random block effects are incorporated into the model. (See both \citet{stroup2012generalized} and \citet{N2014} for a detailed discussion of this approach). An alternative analogous model is the negative binomial model which can be expressed as a Poisson--Gamma mixture model which is often used, for example, in the analysis of next generation sequencing data. In the Poisson--Gamma mixture model, $y_{ij} | \lambda_{ij} v_{ij} \sim \textrm{Poisson}(\lambda_{ij} v_{ij})$ where $v_{ij} \sim \Gamma(1/\phi, \phi)$ for scale parameter $\phi$. %In contrast,  %subtly different to the 

When searching for optimal designs, we need to specify a criterion which describes the relative amount of (usually) treatment information that is available from a design to achieve the objectives of the experiment. Many of the commonly used criteria are based on properties of the Fisher information matrix. Here we focus our attention on finding designs that estimate contrasts of the fixed treatment effects as efficiently as possible, while ensuring that the random effects remain estimable. We then derive these objective functions for the Poisson GLMM for experimental designs with an unstructured treatment factor and a single block factor.

\subsection{Optimality Criteria}
\label{sub:GLMMs-OptCrit}

We consider two optimality criteria: $D_A$--optimality, or generalised $D$--optimality, and $C$--optimality for fixed effects. Our implementation of both of these criteria depends on properties of the partitioned Fisher information matrix in (\ref{eq:FIM-glmm}). 

\citet{ADT2007} describe a $D_A$ optimal design as the design that minimises the determinant $B^{\T} M(\xi,\bs{\beta},\sigma,\sigma_b)^{-1} B$, where $B$ is a set of linear contrasts of the model parameters. We define the $D_A$--optimal design, $\xi_{D_A}^*$, over a class of competing designs, $\mathfrak{X}$, as
\[\xi_{D_A}^* = \textrm{arg} \min_{\xi \in \mathfrak{X}} \det\{B^{\T} M(\xi,\bs{\beta},\sigma,\sigma_b)^{-1} B \},\]
where $\det(\cdot)$ denotes the determinant.

The $C$--optimality criterion is a modification of the $A$--optimality criterion. \citet{ADT2007} define an $A$--optimal design as the design that minimises the trace of the inverse of the Fisher information matrix over $\mathfrak{X}$. That is, the $A$--optimal design is the design, $\xi_A^*$, that is defined as 
\[ \xi_A^* = \textrm{arg} \min_{\xi \in \mathfrak{X}} \tr\{M(\xi,\bs{\beta},\sigma,\sigma_b)^{-1}\}.\]
\citet{ADT2007} then define the $C$--optimal design, $\xi_C^*$, as the design 
 \[\xi_C^* = \textrm{arg} \min_{\xi \in \mathfrak{X}} \tr\{B^{\T} M(\xi,\bs{\beta},\sigma,\sigma_b)^{-1} B \}.\]

We now derive the expression for the $D_A$-- and $C$--optimality objective functions for the estimation of linear combinations of the treatment effects in the model in (\ref{eq:bd-glmm}).

\subsection{Objective Functions}
\label{sub:GLMMs-ObjFn}

Since contrasts of the fixed treatment effects are of interest, let the contrasts in $B$ be linear combinations of the entries in $\bs{\beta}$. In particular, we would like to estimate a set of orthogonal contrasts that form a basis for the degrees of freedom for treatment effects, so that the marginal information matrix is given by
\begin{equation}
\label{eq:bd-glmm-A}
B^{\T} M^{\textrm{marg}}_{\mb \beta}(\xi,\bs{\beta},\sigma,\sigma_b)^{-1} B = B^{\T} \left\{M_{11} - M_{12} (M_{22})^{-1} M_{21}\right\}^{-1} B
\end{equation}
where $M_{11} = X^\T WX$, $M_{12} = M_{21}^\T = X^\T WZ$, $M_{22} = Z^\T WZ+G^{-1}$.

In the case of a Poisson GLMM with a single blocking factor, it follows from the pseudo-likelihood estimating equations in (\ref{eq:pseudoLik-glmm}) that $D = \diag(\lambda_1^{-1},\ldots,\lambda_t^{-1})$, $V_{\lambda}^{1/2} = \diag(\lambda_1^{1/2},\ldots,\lambda_t^{1/2})$, $A = \diag[1/a(\phi)] = I_n$ and, hence, the weight matrix $W = \diag(\lambda_{R(i,j)})$. For a block design with $b$ blocks of size $k$, the block design matrix \ $Z=(I_b\otimes \bs{j}_k | I_{bk})$ and the diagonal covariance matrix corresponding to the random effects assuming $\textrm{cov}(u_i, e_{ij})=0$ for all $i$ and $j$, is
\[\textrm{var}(\bs{q}) = \textrm{var}
\left[
\begin{array}{c}
	\mathbf{u}  \\
	\bs{e}  \\
\end{array}\right]
= G =
\left[
\begin{array}{cc}
	\su2 I_b & \bs{0} \\
	\bs{0}  & \sigma^2I_{bk} \\
\end{array}\right].\]
Substituting these matrix results into $M_{22}$ gives
\begin{align*}
M_{22} &= (I_b\otimes \bs{j}_k | I_{bk})^{\T} \diag(\lambda_{R(i,j)}) (I_b\otimes \bs{j}_k | I_{bk}) +
\left[
\begin{array}{cc}
(1/\su2)I_b & \bs{0} \\
  \bs{0}   & (1/\s2)I_{bk} 
\end{array}
\right]
\end{align*}
Applying the results on the inverse of a sum \citep{henderson1981deriving} to $M_{22}$, i.e. 
\[M_{22} = (G^{-1}+Z^{\T}WZ)^{-1} = G - GZ^{\T}(W^{-1}+ZGZ^{\T})^{-1}ZG,\]
and the fact that $W^{-1}+ZGZ^{\T}$ is block diagonal with the sub-matrix corresponding to the \ith block given by
\[(W^{-1}+ZGZ^{\T})_i = \diag\left(\sigma^2 + \lambda^{-1}_{R(i,j)}\right) + \sigma^2_b \bs{j}_k \bs{j}_k^{\T},\]	
we obtain
\[(W^{-1}+ZGZ^{\T})_i^{-1} = \diag\left(\frac{1}{\sigma^2 + \lambda^{-1}_{R(i,j)}}\right) + \frac{\bs{\ell}_i \bs{\ell}_i^{\T}}{\sigma^2_b \left\{1+(\bs{\ell}^{1/2}_i)^{\T} \bs{\ell}_i^{1/2}\right\}}
,\]
where
\[\bs{\ell}_i = \sigma^2_b\left[ \frac{1}{\sigma^2 + \lambda^{-1}_{R(i,1)}}, \cdots, \frac{1}{\sigma^2 + \lambda^{-1}_{R(i,k)}}  \right].\]
It follows that $M^{\textrm{marg}}_{\bs{\beta}}(\xi,\bs{\beta},\sigma,\sigma_b)$ is block diagonal with the \ith sub-matrix, $M^{\textrm{marg}}_{\bs{\beta}}(\xi,\bs{\beta},\sigma,\sigma_b)_i =  X_i^{\T}\Omega_i X_i$, where $X_i$ contains the rows if the design matrix corresponding to block $i$ and 
\[ \Omega_i =  \diag\left(\frac{1}{\sigma^2 +\lambda^{-1}_{R(i,j)}} \right) 
- \frac{\bs{\ell}_i \bs{\ell}_i^{\T}}{\sigma^2_b\left\{1+(\bs{\ell}_i^{1/2})^{\T}\bs{\ell}_i^{1/2}\right\}}.\]

Since $B$ contains only contrasts of the fixed effects, the expression for $B^{\T} M(\xi,\bs{\beta},\sigma,\sigma_b)^{-1} B$ can be expressed in terms of the marginal information matrix for the fixed effects. It follows that the objective function for the $D_A$--optimal design is given by
\[\xi_{D_A}^* = \underset{\xi \in \mathfrak{X}}{\arg \min} \;\det\left\{B^\T \left( \sum_{i=1}^b X_i^{\T} \Omega_i X_i \right)^{-1} B \right\},\]
while the objective function for the $C$--optimal design is
\[\xi_C^* = \underset{\xi \in \mathfrak{X}}{\arg \min} \; \tr\left\{B^\T \left( \sum_{i=1}^b X_i^{\T} \Omega_i X_i \right)^{-1} B \right\}.\]
A full derivation is provided in the Supplementary Material. The following example illustrates the structure of these objective functions.

\begin{example}
	Suppose that we wish to find the optimal arrangement of $t=3$ treatments in $b=2$ blocks of size $k=3$, and will observe a count response that we wish to model by the Poisson GLMM $\eta_{ij} = \alpha + \tau_{R(i,j)} + u_i + e_{ij}$, where $y_{ij}|u_i,e_{ij} \sim \textrm{Poisson}(\exp(\eta_{ij}))$ for $i=1,2,3$, and $j=1,2$.
	
	The components of the Fisher information matrix, defined in (\ref{eq:bd-glmm-A}), are $Z = (I_2 \otimes \bm{j}_3, I_6)$, $W = \diag(\lambda_{R(i,j)})$, and $G = \diag(\sigma_u^2\j_2^\T,\sigma^2\j_6^\T)$. It follows that
\begin{align*}
M_{22} 
&= [\bs{I}_2\otimes \bs{j}_3 | \bs{I}_{6}]^{\T} \diag(\bs{\lambda}_{R(i)}) [\bs{I}_2\otimes \bs{j}_3 | \bs{I}_{6}] +
\left[
\begin{array}{cc}
(1/\su2)I_2 & \bs{0} \\
\bs{0}                                    & (1/\s2)I_{6} 
\end{array}
\right]\\
&=\left[
\begin{array}{c@{\hspace{0.1cm}}c|c@{\hspace{0.1cm}}c@{\hspace{0.1cm}}c|c@{\hspace{0.1cm}}c@{\hspace{0.1cm}}c}
\lambda_{R(1,\cdot)} &                               0 & \lambda_{R(1,1)} & \lambda_{R(1,2)} & \lambda_{R(1,3)} &                          0 &                          0 &                          0 \\
0 & \lambda_{R(2,\cdot)} &                          0 &                          0 &                         0 &  \lambda_{R(2,1)} & \lambda_{R(2,2)} & \lambda_{R(2,3)} \\
\hline
\lambda_{R(1,1)}       &                              0&   \lambda_{R(1,1)} & 0 & 0 & 0 & 0 & 0 \\
\lambda_{R(1,2)}       &                              0&                           0 & \lambda_{R(1,2)} & 0 & 0 & 0 & 0 \\
\lambda_{R(1,3)}       &                              0&                           0 & 0 & \lambda_{R(1,3)} & 0 & 0 & 0 \\
\hline
0                               & \lambda_{R(2,1)}     &                           0 & 0 & 0 & \lambda_{R(2,1)} & 0 & 0 \\
0                               & \lambda_{R(2,2)}     &                           0 & 0 & 0 & 0 & \lambda_{R(2,2)} & 0 \\
0                               & \lambda_{R(2,3)}     &                           0 & 0 & 0 & 0 & 0 & \lambda_{R(2,3)}
\end{array}
\right]\\
& \hspace{1cm} +
\left[
\begin{array}{cc}
(1/\su2)I_2 & \bs{0} \\
\bs{0}     & (1/\sigma^2)I_{6} 
\end{array}
\right]\\
%&=\begin{pmatrix}
%\bs{\lambda}_{R(i,\cdot)}^\delta & (\bs{\lambda}_{R(1)}| \mb{0}_3)^\T & %(\mb{0}_3|\bs{\lambda}_{R(2)})^\T  \\
%\mb{0}_3 & \bs{\lambda}_{R(1)}^\delta & \mb{0}_{3\times 3} \\
%(\mb{0}_3|\bs{\lambda}_{R(2)}) & \mb{0}_{3\times 3} & \bs{\lambda}_{R(2)}^\delta
%\end{pmatrix}
%+
%\begin{pmatrix}
%(1/\su2)I_2 & \bs{0} \\
%\bs{0}     & (1\s2)I_{6} 
%\end{pmatrix}
\end{align*}
where  $\lambda_{R(i,\cdot)} = \sum_{j=1}^{k} \lambda_{R(i,j)}$ and %$\bs{\ell}_i$ is define as in .
\[\bs{\ell}_i = \left[ \frac{\sigma^2_b}{\sigma^2 + \lambda^{-1}_{R(i,1)}}, \frac{\sigma^2_b}{\sigma^2 + \lambda^{-1}_{R(i,2)}}, \frac{\sigma^2_b}{\sigma^2 + \lambda^{-1}_{R(i,3)}}  \right].
\]
The \ith sub-matrix of the marginal information matrix, corresponding to block $i$ in the design, is given by
\begin{align*}
\hspace{2em}&\hspace{-2em}M^{\textrm{marg}}_{\mb \beta}(\xi,\bs{\beta},\sigma,\sigma_b)_i\\
 &= \diag\left(\frac{1}{\sigma^2 + \lambda^{-1}_{R(1,j)}}\right) - \frac{\bs{\ell}_1 \bs{\ell}_1^{\T}}{\sigma^2_b \left(1+(\bs{\ell}^{1/2}_1)^{\T} \bs{\ell}_1^{1/2}\right)}\\
&= \left[ \begin{array}{ccc}
\frac{1}{\sigma^2 + \lambda^{-1}_{R(1,1)}} & 0 & 0 \\ 
0& \frac{1}{\sigma^2 + \lambda^{-1}_{R(1,2)}} & 0 \\ 
0& 0 & \frac{1}{\sigma^2 + \lambda^{-1}_{R(1,3)}}
\end{array} \right] \\
& \hspace{1cm} - \frac{1}{1+\frac{\sigma^2_b}{\sigma^2 + \lambda^{-1}_{R(1,1)}}+\frac{\sigma^2_b}{\sigma^2 + \lambda^{-1}_{R(1,2)}}+\frac{\sigma^2_b}{\sigma^2 + \lambda^{-1}_{R(1,3)}}} \times \\
& \hspace{1cm}  \left[ \begin{array}{ccc}
\frac{\sigma^2_b}{(\sigma^2 + \lambda^{-1}_{R(1,1)})^2} & \frac{\sigma^2_b}{(\sigma^2 + \lambda^{-1}_{R(1,1)})(\sigma^2 + \lambda^{-1}_{R(1,2)})} & \frac{\sigma^2_b}{(\sigma^2 + \lambda^{-1}_{R(1,1)})(\sigma^2 + \lambda^{-1}_{R(1,3)})} \\ 
\frac{\sigma^2_b}{(\sigma^2 + \lambda^{-1}_{R(1,2)})(\sigma^2 + \lambda^{-1}_{R(1,1)})}& \frac{\sigma^2_b}{(\sigma^2 + \lambda^{-1}_{R(1,2)})^2} & \frac{\sigma^2_b}{(\sigma^2 + \lambda^{-1}_{R(1,2)})(\sigma^2 + \lambda^{-1}_{R(1,3)})} \\ 
\frac{\sigma^2_b}{(\sigma^2 + \lambda^{-1}_{R(1,3)})(\sigma^2 + \lambda^{-1}_{R(1,1)})}& \frac{\sigma^2_b}{(\sigma^2 + \lambda^{-1}_{R(1,3)})(\sigma^2 + \lambda^{-1}_{R(1,2)})} & \frac{\sigma^2_b}{(\sigma^2 + \lambda^{-1}_{R(1,3)})^2}
\end{array} \right],
\end{align*}
with the corresponding structure of the second block taking a similar form. 

To investigate the optimal designs that are produced, Table \ref{tab:smalleg} gives the $D_A$--optimal and $C$--optimal designs for a variety of treatment means and and values of $\sigma_b^2$, with $\sigma^2 = 0.25$. We observe that as the size of the block variance increases relative to the treatment means, the optimal design becomes more balanced. For small block variances the effect of the different treatment variances becomes more dominant in determining the optimal design.  
\end{example}

\begin{table}
	\caption{$D_A$- and $C$-optimal designs for three treatments with expected counts $(\lambda_1, \lambda_2, \lambda_3)=$ (1,1,1), (1,1,2), (1,2,4) or (1,4,16), block variance $\sigma_b^2 = 0.016$, $0.25$, or $4$ and $\sigma^2 = 0.25$. The relative efficiencies of the randomised complete block design are given in the final column}\label{tab:smalleg}
	\begin{center}
		\begin{tabular}{lllllclcl}
			\hline
			& & & & & \multicolumn{2}{c}{$D_A$--optimality} & \multicolumn{2}{c}{$C$--optimality}\\
			
			& & & & & Optimal & \multicolumn{1}{c}{BIBD} & Optimal & \multicolumn{1}{c}{BIBD}\\
			$\lambda_1$ & $\lambda_2$ & $\lambda_3$ & $\sigma_b^2$ & $\sigma^2$ & design & \multicolumn{1}{c}{Efficiency} & design & \multicolumn{1}{c}{Efficiency}\\
			\hline
			1 & 1 &  1 & 0.016 & 0.25 & (1,2,3), (1,2,3) & \quad 1 & (1,2,3), (1,2,3) & \quad 1\\
			1 & 1 &  2 & 0.016 & 0.25 & (1,2,3), (1,2,3) & \quad 1 & (1,1,2), (1,2,3) & \quad 0.988\\
			1 & 2 &  4 & 0.016 & 0.25 & (1,2,3), (1,2,3) & \quad 1 &  (1,1,2), (1,2,3) & \quad 0.919\\
			1 & 4 & 16 & 0.016 & 0.25 & (1,2,3), (1,2,3) & \quad 1 & (1,1,2), (1,2,3) & \quad 0.851\\
			1 & 1 &  1 & 0.25  & 0.25 & (1,2,3), (1,2,3) & \quad 1 & (1,2,3), (1,2,3) & \quad 1\\
			1 & 1 &  2 & 0.25  & 0.25 & (1,2,3), (1,2,3) & \quad 1 & (1,2,3), (1,2,3) & \quad 1\\
			1 & 2 &  4 & 0.25  & 0.25 & (1,2,3), (1,2,3) & \quad 1 & (1,1,2), (1,2,3) & \quad 0.990\\
			1 & 4 & 16 & 0.25  & 0.25 & (1,2,3), (1,2,3) & \quad 1 & (1,1,2), (1,2,3) & \quad 0.923\\
			1 & 1 &  1 &    4  & 0.25 & (1,2,3), (1,2,3) & \quad 1 & (1,2,3), (1,2,3) & \quad 1\\
			1 & 1 &  2 &    4  & 0.25 & (1,2,3), (1,2,3) & \quad 1 & (1,2,3), (1,2,3) & \quad 1\\
			1 & 2 &  4 &    4  & 0.25 & (1,2,3), (1,2,3) & \quad 1 & (1,2,3), (1,2,3) & \quad 1\\
			1 & 4 & 16 &    4  & 0.25 & (1,2,3), (1,2,3) & \quad 1 & (1,2,3), (1,2,3) & \quad 1\\
			\hline
		\end{tabular}
	\end{center}
\end{table}

\section{Locally optimal block designs using simulated annealing}
\label{sec:simulated-annealing}

Generating \DaOpt{} or \COpt{} block designs requires an iterative 
search algorithm for which we have elected to use simulated annealing (SA) \citep{KGV1983}. Since the design criterion to be minimised for both $D_\textrm{A}$-- and \COpt{}ity is the generalised variance of $B\bs{\beta}$, which for count data is functionally dependent on the treatment group means, we constrain the SA algorithm to search for locally optimal designs, i.e. designs that are optimal for a set of point priors for the expected treatment counts, $\lambda_{h}$, $h=1,\ldots,t$, and the variance components between blocks, $\su2$, and residuals, $\s2$. %Since simulated annealing searches for candidate designs that minimise the objective function, and avoid local minima by occasionally moving to a design with a higher value of the objective function, we set the objective function to minimise $-|\mbC_{11}|$, i.e. the negative of the determinant of the information matrix.

Three key inputs are required by the SA algorithm: a starting design, an objective function and candidate generator procedure, which we now discuss.

The SA algorithm is initialised with a starting design, $D_0$, generated by randomly assigning treatments to blocks with objective function value $O(D_0)$. At each iteration a new design, $D_i$, is generated by random exchanges of treatments in randomly selected experimental units in design $D_{i-1}$ at the previous iteration, where $D_{i-1}=D_0$ at the first iteration. Since the SA algorithm searches for candidate designs which minimise the objective function, $D_{i}$ always replaces $D_{i-1}$ if $O(D_{i})<O(D_{i-1})$, and has a small probability of replacing $D_{i-1}$ even it is a slightly worse design. The acceptance probability of a worse design depends on the so-called temperature of the algorithm, which is initially set high to enable the algorithm to escape local optima in early iterations. As the iterations continue the temperature gradually cools, and with it the probability of accepting worse designs. In this way, the algorithm converges to the global optimum within the constrained set of competing designs. 

As discussed in section~\ref{sub:GLMMs-ObjFn}, we consider two objective functions based on the \DaOpt{}ity and \COpt{}ity criteria, both defined in \citet{ADT2007}. The goal is search the space of candidate designs that minimise these functions.

In contrast to block designs from the classical setting, where optimal efficiency is achieved by allocating treatments as equally as possible among blocks, optimal designs based on responses drawn from Poisson distributions have treatment replication inversely proportional to their treatment means. Our selection of a starting design, therefore, makes no assumption of equal replication or balance. Consequently, a candidate design generating procedure which makes random exchanges of treatments between blocks is unsatisfactory. Instead, we propose starting with a random design and then substitute the treatment assigned to a randomly chosen experimental unit in the design with a randomly chosen treatment from the treatment set. 

Our preliminary testing of this strategy showed that, for some sets of design parameters (i.e. number of treatments, blocks and block size) and point priors, the SA algorithm converged very slowly and sometimes would get caught in local minima, even for reasonably high initial temperatures. To overcome these limitations, our candidate design generating procedure includes an option for $m \geqslant 1$ substitutions to be made at each iteration. A vector of probabilities $P=\{(p_1,\ldots,p_m): p_1\geq \cdots \geq p_m\textrm{ and } p_1+\cdots+p_m=1\},$ where $p_m$ denotes the probability that $m$ experimental units will have treatment substitutions at a given iteration.

The SA algorithm for performing the optimisation described above is implemented in the \texttt{designGLMM} \texttt{R} package which is available from the Comprehensive R Archive Network. 

We now consider two examples where we find optimal block designs for experiments in which responses are drawn from a Poisson distribution using our SA algorithm.

\section{Examples}
\label{sec:example}

\subsection{Differential striatal gene expression between two strains of mouse}

We consider a comparative experiment to assess the level of differential striatal gene expression between two mice strains using the Illumina GAIIx next-generation sequencing (NGS) platform \citep{BWH2011}. cDNA, copied from amplified RNA isolated from cells in the striatum of twenty-one mice -- ten from the C57BL/6J strain (strain 1) and eleven from the DBA/2J strain (strain 2) -- was loaded into individual lanes (plots) of three flow cells (blocks) for sequencing. The design used by \cite{BWH2011} comprised two flow cells with three replicates of strain 1 and four replicates of strain 2, and a third flow cell with four replicates of strain 1 and three replicates of strain 2. The question that we wish to answer is whether this design, or a different design with the same number of samples, is optimal for the estimation of strain effects.

Table \ref{tab bottomly data} presents counts from four of the 36536 identified genes (labelled A = ENSMUSG00000046994, B = ENSMUSG00000039967, C = ENSMUSG00000050141 and D = ENSMUSG00000033826) in this experiment, selected to represent the different effect sizes observed across the entire data set (see ReCount resource \citep{FLL2011}). A per-gene GLMM of the form presented in (\ref{eq:bd-glmm}) was fitted to the count data yielding the parameter estimates shown in Table~\ref{tab:NGS-param-est}. We now use the effect sizes obtained from these estimates as point priors in searching for optimal designs for the estimation of the strain effect.

Table~\ref{tab:NGS-param-est} shows that the size of the strain effect in genes A and B are quite small, with the relative abundances being approximately equal to 1. However, these genes do differ when we consider the ratio of the between flow cell variation to the within flow cell variation (i.e. $\su2/\s2$) of each. For gene A the between flow cell variation is 0.4 times that within cells, while for gene B this variance ratio is an order of magnitude larger. The \COpt{} design based on the point priors estimated from the gene A data consists of three flow cells, each comprising four replicates of strain 1 and three replicates of strain 2, while the \COpt{} design based on the point priors estimated from the gene B data consists of three flow cells, each comprising three replicates of strain 1 and four replicates of strain 2.

In contrast, the strain effect is very large for genes C and D with 2000 more copies of gene C in strain 1 than strain 2 and, conversely, almost 28 times the number of copies of gene D in strain 2 than strain 1. For both of these genes the magnitude of the variation between flow cells is comparable with that for genes A and B, however the within flow cell variation for genes C and D appears negligible.  The \COpt{} design based on the point priors estimated from the gene C data consists of three flow cells, each comprising one replicate of strain 1 and replicates of strain 2, while  the \COpt{} design based on the point priors estimated from the gene D data consists of three flow cells, each comprising five replicates of strain 1 and two replicates of strain 2. Neither of these designs, nor those identified as optimal for genes A and B, is the same as the design used by \cite{BWH2011}.

For each of genes A -- D we now consider the relative performances of eight alternative designs, $D_1$ -- $D_8$ shown in Table~\ref{tab:bottomly}, in which the twenty-one striatum cDNA samples are arranged in three blocks (flow cells), with seven samples per block. The treatments (strains) assigned to each flow cell are denoted by $1^{r_1}2^{r_2}$, where $r_h$ denotes the number of replicates of strain $h$, $h=1,2$. Designs $D_3$, $D_6$, $D_8$ and $D_2$ are the \COpt{} designs given above for genes A to D, respectively. Designs $D_4$, with two blocks containing $1^42^3$ and one block containing $1^32^4$, and $D_5$, with two blocks each containing $1^32^4$ and one block containing $1^32^4$, would be considered optimal and isomorphic under unit-treatment additivity. 

\begin{figure}
  % trim=left bottom right top
  \includegraphics[trim={0cm 0cm 0cm 1.4cm},clip, width=1\textwidth]{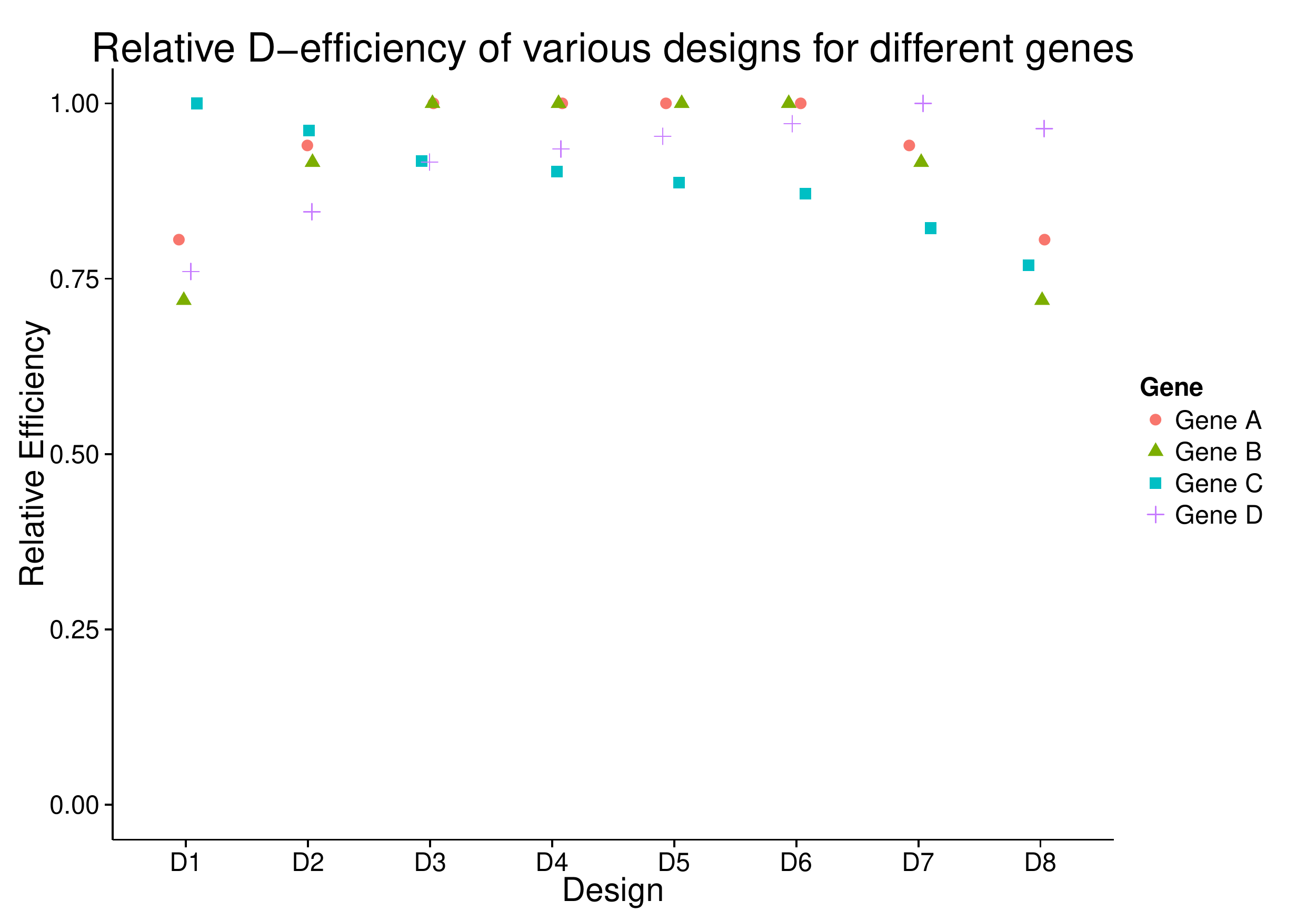}\\
  \caption{Per-gene relative efficiencies of eight designs, $D_1$ -- $D_8$, defined in Table~\ref{tab:bottomly} based on point priors of four genes, A -- D, given in Table~\ref{tab:NGS-param-est}.}
  \label{fig:DesignEffs}
\end{figure}

Figure~\ref{fig:DesignEffs} shows the per-gene relative efficiencies of designs $D_1$ -- $D_8$ using the point priors in Table~\ref{tab:bottomly}, where here we define  relative efficiency of design $D_i$ as $O_g(D_i)/\max\{O_g(D_1),\ldots,O_g(D_8)\}$, i.e. the ratio of the value of the objective function, based on \COpt{ity}, for design $D_i$ relative to the largest value of the objective function across all eight designs, based on the point priors of gene $g$, $g=$ A, B, C, D. Figure~\ref{fig:DesignEffs} shows that designs $D_4$ and $D_5$ are optimal for genes A and B which each have a negligible strain effect. Note that because these design are near-balanced they are also \DaOpt{}. These designs are not optimal, however, for genes C and D, where the strain effects are quite large. Indeed, the larger the strain effect, the more substantial the loss in efficiency.

\begin{table}
 	\caption{Gene counts for four selected genes from two strains of mice: C57BL/6J (1) and DBA/2J (2). The columns within each subtable correspond to the seven lanes into which individual cDNA samples were loaded within a flow cell.}
    \label{tab bottomly data}
    {\fontsize{11}{13.2}\selectfont
	{\renewcommand{\arraystretch}{1.1}
	\begin{center}
		\begin{tabular}{ccrrrrrrr}
           \hline
           && \multicolumn{7}{c}{Strain}\\
           \cline{3-9}
            Flow cell & Gene$^\dagger$ &    1 &   1 &   1 &   2 &   2 &   2 &   2 \\ \hline
               1 &    A &  132 & 134 & 140 & 112 & 134 & 100 & 115 \\
               1 &    B &  794 & 922 & 606 & 507 & 688 & 510 & 659 \\
               1 &    C &   34 &  59 &  52 &   1 &   0 &   0 &   1 \\
               1 &    D &   10 &  12 &   7 &   9 &  38 &  29 &  19 \\ \hline
           \\ %\vspace{-1mm}
           \hline
           && \multicolumn{7}{c}{Strain}\\
           \cline{3-9}
           Flow cell & Gene$^\dagger$ &   1 &   1 &   1 &   1 &    2 &   2 &   2 \\ \hline
              2 &  A   & 101 &  68 &  64 & 102 &  132 & 139 & 110 \\
              2 &  B   & 758 & 722 & 731 & 803 & 1080 & 614 & 961 \\
              2 &  C   &  43 &  29 &  30 &  31 &    0 &   0 &   1 \\
              2 &  D   &  41 &   1 &  12 &   3 &    2 &  33 &  61 \\ \hline
           \\ %\vspace{-1mm}
           \hline
           && \multicolumn{7}{c}{Strain}\\
           \cline{3-9}
           Flow cell & Gene$^\dagger$ &    1 &    1 &    1 &    2 &    2 &    2 &    2 \\ \hline
              3 &  A   &  174 &  194 &  194 &  146 &  155 &  157 &  128 \\
              3 &  B   & 1169 & 1353 & 1343 & 1359 & 1437 & 1426 & 1512 \\
              3 &  C   &   64 &   41 &   56 &    6 &    1 &    1 &    1 \\
              3 &  D   &   18 &    5 &    5 &   35 &   50 &   31 &   45 \\ \hline
           \multicolumn{9}{l}{$^\dagger$Genes: A = ENSMUSG00000046994; B = ENSMUSG00000039967}\\
           \multicolumn{9}{l}{\phantom{$^\dagger$Genes: }C = ENSMUSG00000050141; D = ENSMUSG00000033826}
		\end{tabular}
	\end{center} } \par}
\end{table}

%\begin{table}
%	\caption{Data for four selected genes from mouse expression data. Strain 1 corresponds to C57BL/6J and Strain 2 corresponds to DBA/2J}\label{tab bottomly data}
%	\begin{center}
%		\begin{tabular}{lrrrrrrrrrrrrrrrrrrrrr}
%			\hline
%			Strain&1&1&1&1&1&1&1&1&1&1&2&2&2&2&2&2&2&2&2&2&2\\
%			Chip&1&1&1&2&2&2&2&3&3&3&1&1&1&1&2&2&2&3&3&3&3\\
%			\hline
%			ENSMUSG00000046994&132&134&140&101&68&64&102&174&194&194&112&134&100&115&132&139&110&146&155&157&128\\
%			ENSMUSG00000039967&794&922&606&758&722&731&803&1169&1353&1343&507&688&510&659&1080&614&961&1359&1437&1426&1512\\
%			ENSMUSG00000050141&34&59&52&43&29&30&31&64&41&56&1&0&0&1&0&0&1&6&1&1&1\\
%			ENSMUSG0000003382610&12&7&9&1&12&3&2&5&5&35&38&29&19&41&33&61&18&50&31&45\\
%			ENSMUSG00000074233&1&2&1&1&2&2&0&3&2&3&0&2&2&1&2&0&2&2&1&4&3\\
%			\hline
%		\end{tabular}
%	\end{center}
%\end{table}

\begin{table}
	\caption{Parameter estimates on the link scale ($\alpha,\tau_1,\sigma,\sigma_u$) and the response scale ($\lambda_1,\lambda_2$) from fitting a per-gene Poisson GLMM to gene counts.}
    \label{tab:NGS-param-est}
    {\fontsize{11}{13.2}\selectfont
    \setlength{\tabcolsep}{5pt}
	{\renewcommand{\arraystretch}{1.2}
	\begin{center}
		\begin{tabular}{crrrrp{0cm}rr}
			\hline
		    & \multicolumn{4}{c}{Link scale} && \multicolumn{2}{c}{Response scale}\\
		    \cline{2-5}\cline{7-8}
			Gene & \multicolumn{1}{c}{$\alpha$} & \multicolumn{1}{c}{$\tau_1$} & \multicolumn{1}{c}{$\sigma$} & \multicolumn{1}{c}{$\sigma_u$}&& \multicolumn{1}{c}{$\lambda_1$} 
			& \multicolumn{1}{c}{$\lambda_2$}\\
			\hline
			A & 4.85767 &  0.00050 & 0.20104 & 0.12874 && 128.66  & 128.79 \\
			B & 6.81209 & -0.00001 & 0.13382 & 0.27905 && 908.77  & 908.76 \\
			C & 3.78631 & -3.73949 & 0.00000 & 0.19885 && 1855.30 &   1.05 \\
			D & 1.87168 &  1.66639 & 0.00002 & 0.26546 && 1.23    &  34.40 \\
			\hline
		\end{tabular}
	\end{center}}\par}
\end{table}

%\begin{table}
%	\caption{Parameter estimates for four selected genes from mouse expression data}\label{tab param}
%	\begin{center}
%		\begin{tabular}{crrrr}
%			\hline
%			Gene & $\beta_0$ & $\beta_1$ & $\sigma$ & $\sigma_b$\\
%			\hline
%			ENSMUSG00000046994&4.85767&0.00050&0.20104&0.12874\\
%			ENSMUSG00000039967&6.81209&-0.00001&0.13382&0.27905\\
%			ENSMUSG00000050141&3.78631&-3.73949&0.00000&0.19885\\
%			ENSMUSG00000033826&1.87168&1.66639&0.00002&0.26546\\
%			ENSMUSG00000074233&0.51890&0.00008&0.00002&0.20012\\
%			\hline
%		\end{tabular}
%	\end{center}
%\end{table}

\begin{table}
	\caption{Eight block designs for $t=2$ strains arranged in $b=3$ blocks of size $k=7$. Treatments within a block are denoted by $1^{r_1}2^{r_2}$, where $r_h$ denotes the number of replicates of strain $h$, $h=1,2$. The blocks in designs $D_4$ and $D_5$ each have two different combinations of replicates of strains 1 and 2, indicated by the multiplier $c$ in $c\times 1^{r_1}2^{r_2}$, $c=1,2$.}
	\label{tab:bottomly}
	{\renewcommand{\arraystretch}{1.2}
		\begin{center}
		\begin{tabular}{cccccccc}
			\hline
			            \multicolumn{8}{c}{Design} \\
            $D_1$             & $D_2$ & $D_3$ & $D_4$ & $D_5$    & $D_6$   & $D_7$     & $D_8$ \\
			\hline
			$1^62^1$ & $1^52^2$ & $1^42^3$ & $2\times 1^42^3$ & $2\times 1^32^4$ & $1^32^4$ & $1^22^5$ & $1^12^6$  \\
			         &          & & $1\times 1^32^4$ & $1\times 1^42^3$ &  &  &   \\ \hline
		\end{tabular}
	\end{center}}
\end{table}

%\begin{table}
%	\caption{D--optimal designs for mouse expression data based on parameters estimated from (a) gene ENSMUSG00000046994, (b) gene ENSMUSG00000039967, (c) gene ENSMUSG00000050141, and (d) gene ENSMUSG00000033826}\label{tab bottomly}
%	\subfloat[(a)]{
%		\begin{tabular}{c}
%			\hline
%			1    1    1    1    2    2    2\\
%			1    1    1    1    2    2    2\\
%			1    1    1    1    2    2    2\\
%			\hline
%		\end{tabular}
%		}
%	\subfloat[(b)]{
%		\begin{tabular}{c}
%			\hline
%			1    1    1    2    2    2    2\\
%			1    1    1    2    2    2    2\\
%			1    1    1    2    2    2    2\\
%			\hline
%		\end{tabular}
%		}
%	\subfloat[(c)]{
%		\begin{tabular}{c}
%			\hline
%			1    2    2    2    2    2    2\\
%			1    2    2    2    2    2    2\\
%			1    2    2    2    2    2    2\\
%			\hline
%			\end{tabular}
%			}
%	\subfloat[(d)]{
%		\begin{tabular}{c}
%			\hline
%			1    1    1    1    1    2    2\\
%			1    1    1    1    1    2    2\\
%			1    1    1    1    1    2    2\\
%			\hline
%		\end{tabular}
%		}
%\end{table}

\subsection{Begging behaviour of nestling barn owls}

\cite{RB2007} investigated the begging behaviour of nestling barn owls. They recorded the number of begging vocalisations, or calls, made by an offspring to its parent in the 15 minutes prior to the parent owl's arrival at the nest. Of interest were the treatment factors gender (of the parent) and satiety (food-deprived and food-satiated juvenile). See \cite{ZIW2009} for a detailed discussion of these data. 

Suppose that in a future similar study the researchers want to investigate 15 barn owl broods (blocks) each comprising 10 nestlings, what would be the optimal design? Treating the four combinations of gender and satiety as four levels of a single treatment factor, we fitted a Poisson GLMM to the data given in \citep{RB2007} to obtain estimates of the requisite point priors.  From this analysis we found that the mean number of calls for Deprived Females was $\lambda_1 = 1.33$, for Deprived Males was $\lambda_2 = 1.36$, for Satiated Females was $\lambda_3 = 0.44$ and for Satiated Males was $\lambda_4 = 0.54$. The between nest standard deviation was $\sigma_u = 1.11$ and the within nest excess variation was $\sigma = 0.47$. The \COpt{} design for this experiment is the one in which all 15 broods comprise the treatment allocation $1^32^33^24^2$ to nestlings. On the other hand, the classically optimal design would consist of two broods each with the treatment allocations $1^32^33^24^2$, $1^32^23^34^2$, $1^32^23^24^3$, $1^22^33^34^2$, $1^22^33^24^3$, and $1^22^23^34^3$, with the remaining three broods being selected from these six combinations such that the treatments are as balanced as possible ($1^32^33^24^2$, $1^22^23^34^3$, and $1^22^33^24^3$, for instance). 

\section{Discussion}
\label{s:discuss}

In this paper, we find optimal designs for Poisson regression with a single unstructured treatment factor and a single variable that creates blocks of equal size. The methods discussed here are implemented in the \texttt{R} statistical software package \texttt{designGLMM}, which is available on the Comprehensive R Archive Network (cran.r-project.org) under a GPL3 license. We observe that for experiments where the treatment means are sufficiently different, and the block effect is not dominant, the optimal designs differ from those for linear models. This is because in a non--linear setting the Fisher information matrix, and hence the optimal designs, depend on the values of the model parameters. 

The optimal designs for generalised linear mixed models depend on the functional form of the model and the values of the model parameters. In this paper, we have used a Poisson--Lognormal model, as discussed in \citet{stroup2012generalized} and \citet{N2014}. Many other model configurations are available for modelling count data, most notably the negative binomial model (\citet{lawless1987negative}). \cite{hilbe2011negative} presents yet other possibilities, including alternate mean--variance relationships and hurdle models. Use of these alternate modelling approaches may yield optimal designs for the estimation of treatment effects which differ from those based on the Poisson GLMM. 

The objective functions considered in this paper assume that the treatment effects are of primary interest. Specifically, a set of linear combinations of treatment effects are to be estimated with as small variance as possible, while being distinguishable from block effects. In some experiments, researchers are also interested in estimating the block effects efficiently, which may give rise to different optimal designs.

In Example 1 of Section~\ref{sec:example}, we considered the optimal design of a NGS experiment in which seven samples were placed onto individual lanes of three different flow cells. In generating optimal designs for this experiment, we considered only the variability between chips, and not between lanes. \citet{AD2010} suggest that variation between lanes should also be a consideration. Furthermore, some NGS experiments use a process called \emph{barcoding} to place multiple samples onto a single lane. This would suggest that more complex design structures, such as row--column designs, may be appropriate. 

Additional complications arise from the design of NGS experiments. For instance, \citet[Eq.3]{AD2010} use an offset term, such as $\log(c_{ij})$, to normalise the number of reads per lane, and is common practise in the modelling of NGS data (see, for example Mortazavi et al. 2008). 

We are currently looking at how we can address some of these issues in the optimal design of NGS experiments. Other areas which require further investigation include incorporating prior distributions for each of the model parameters to develop Bayesian optimal designs, and the investigation of alternative search algorithms that may be more efficient than simulated annealing in finding optimal designs. 

%  The \backmatter command formats the subsequent headings so that they
%  are in the journal style.  Please keep this command in your document
%  in this position, right after the final section of the main part of 
%  the paper and right before the Acknowledgements, Supplementary Materials,
%  and References sections. 

%  This section is optional.  Here is where you will want to cite
%  grants, people who helped with the paper, etc.  But keep it short!

%\section*{Acknowledgements}

%The authors thank Professor A. Sen for some helpful suggestions,
%Dr C. R. Rangarajan for a critical reading of the original version of the
%paper, and an anonymous referee for very useful comments that improved
%the presentation of the paper.\vspace*{-8pt}

%  If your paper refers to supplementary web material, then you MUST
%  include this section!!  See Instructions for Authors at the journal
%  website http://www.biometrics.tibs.org

\section*{Supplementary Materials}

Web Appendix A, referenced in Section~\ref{sec:GLMMs-CorrCount}, is available with
this paper at the Biometrics website on Wiley Online
Library.\vspace*{-8pt}

%  Here, we create the bibliographic entries manually, following the
%  journal style.  If you use this method or use natbib, PLEASE PAY
%  CAREFUL ATTENTION TO THE BIBLIOGRAPHIC STYLE IN A RECENT ISSUE OF
%  THE JOURNAL AND FOLLOW IT!  Failure to follow stylistic conventions
%  just lengthens the time spend copyediting your paper and hence its
%  position in the publication queue should it be accepted.

%  We greatly prefer that you incorporate the references for your
%  article into the body of the article as we have done here 
%  (you can use natbib or not as you choose) than use BiBTeX,
%  so that your article is self-contained in one file.
%  If you do use BiBTeX, please use the .bst file that comes with 
%  the distribution.  In this case, replace the thebibliography
%  environment below by 
%

\def\newblock{\hskip .11em plus .33em minus .07em}

\appendix

\section{Full derivation of $M^{\textrm{marg}}_{\pmb \beta}(\xi,\pmb{\beta},\sigma,\sigma_b)$}

In this section, we present a full derivation for the marginal information matrix for the estimation of a set of contrasts of the fixed parameters $B\pmb{\beta}$ for a Poisson GLMM. The marginal information matrix will is given by
\begin{equation*}
B^T M^{\textrm{marg}}_{\pmb{\beta}}(\xi,\pmb{\beta},\sigma,\sigma_b)^{-1} B = B^T \left[M_{11} - M_{12} \{M_{22}\}^{-1} M_{21}\right]^{-1} B
\end{equation*}
where $M_{11} = X^\T WX$, $M_{12} = M_{21}^\T = X^\T WZ$, and $M_{22} = Z^\T WZ+G^{-1}$. In this formulation, $W =( DV_{\lambda}^{1/2}AV_{\lambda}^{1/2}D)^{-1}$, and $G$ is the covariance matrix of random effects. For Poisson regression, we have
\begin{align*}
D &= diag\left[\frac{\partial g(\pmb{\lambda}|\pmb{bu})}{\partial \pmb{\lambda}}\right] = diag\left[\frac{\partial \log(\pmb{\lambda})}{\partial \pmb{\lambda}}\right] = diag[\lambda_1^{-1},\lambda_2^{-1},\ldots,\lambda_t^{-1}] , \\
V_{\lambda}^{1/2} &= diag\left[\left(\frac{\partial^2 b(\theta)}{\partial \theta^2}\right)^{1/2}\right] = diag\left[\left(\frac{\partial^2 \exp(\pmb{\eta})}{\partial \pmb{\eta}^2}\right)^{1/2}\right] = diag[(\lambda_1,\lambda_2,\ldots,\lambda_t)^{1/2}],\\
A &= diag[1/a(\phi)] = I_N,
\end{align*}
and hence $W = diag[\lambda_{R(i,j)}]$. If we assume that $\textrm{cov}(u_i, e_{ij})=0$ for all $i,j$ then
\[\textrm{Var}(\pmb{q}) = \textrm{var}
\left[
\begin{array}{c}
\pmb{u}  \\
\pmb{e}  \\
\end{array}\right]
= G =  \left[
\begin{array}{cc}
\sigma^2_b I_b & \pmb{0} \\
\pmb{0}                      & \sigma^2 I_{bk} \\
\end{array}\right].\]
We can use this information to simplify the expression for $(M_{22})^{-1}$ so that it does not require matrix inversion. This will improve computation times for the simulated annealing algorithm.

If we consider a block design with $b$ blocks of size $k$. We then have that the block design matrix, $Z=[I_b\otimes \pmb{j}_k | I_{bk}]$, weight matrix $W=diag(\lambda_{R(i,j)})$ and diagonal covariance matrix corresponding to the random effects $G$. Then
\begin{align*}
M_{22} &= Z^TWZ + G^{-1} \\
&= [\pmb{I}_b\otimes \pmb{j}_k | \pmb{I}_{bk}]^Tdiag(\lambda_{R(i,j)}) [\pmb{I}_b\otimes \pmb{j}_k | \pmb{I}_{bk}] +
\left[
\begin{array}{cc}
\frac{1}{\sigma^2_b}\pmb{I}_b & \pmb{0} \\
\pmb{0}                                    & \frac{1}{\sigma^2}\pmb{I}_{bk} 
\end{array}
\right]
\end{align*}

Using the inverse sum of matrices result of \citet{henderson1981deriving} that
\[(H+JKL)^{-1} = H^{-1} - H^{-1}J(K^{-1}+LH^{-1}J)^{-1}LH^{-1},\]
we obtain
\[(G^{-1}+Z^TWZ)^{-1} = G - GZ^T(W^{-1}+ZGZ^T)^{-1}ZG.\]
Now
\begin{align*}
W^{-1}+ZGZ^T &= diag\left(\lambda^{-1}_{R(i,j)}\right) + \left[I_b \otimes \pmb{j}_k | I_{bk}\right] \left[\begin{array}{cc}
\sigma^2_b I_b & \pmb{0}\\
\pmb{0}           & \sigma^2 I_{bk}
\end{array} \right]
\left[\begin{array}{l}
I_b \otimes \pmb{j}_k^T\\
I_{bk} 
\end{array}\right]\\
&= diag\left(\lambda^{-1}_{R(i,j)}\right) +  \left[\sigma^2_b I_b \otimes \pmb{j}_k \pmb{j}_k^T + \sigma^2 I_{bk}\right]\\
&= diag\left(\sigma^2 + \lambda^{-1}_{R(i,j)}\right) + \sigma^2_b\left[\begin{array}{cccc}
\pmb{j}_k \pmb{j}_k^T & \pmb{0}                      & \cdots & \pmb{0}\\
\pmb{0}                      & \pmb{j}_k \pmb{j}_k^T & \cdots & \pmb{0}\\
\vdots                         & \vdots                         & \ddots & \vdots\\ 
\pmb{0}                      & \pmb{0}                      & \cdots & \pmb{j}_k \pmb{j}_k^T
\end{array}\right]
\end{align*} 

Notice that this matrix is block diagonal, with $b$ blocks of size $k\times k$ with similar structure. We can then express the $(i,i)^{\textrm{th}} $ block as
\[(W^{-1}+ZGZ^T)_{i} = diag\left(\sigma^2 + \lambda^{-1}_{R(i,j)}\right) + \sigma^2_b \pmb{j}_k \pmb{j}_k^T\]
This is of the form $(H+\pmb{a}\pmb{b}^T)$, where $H$ is invertable and square and $\pmb{a}$ and $\pmb{b}$ are column vectors, so we can invert this block using the Sherman-Morrison formula. Then
\[(W^{-1}+ZGZ^T)_{i}^{-1} = diag\left(\frac{1}{\sigma^2 + \lambda^{-1}_{R(i,j)}}\right) - \frac{ diag\left(\frac{1}{\sigma^2 + \lambda^{-1}_{R(i,j)}}\right) \times \sigma^2_b \pmb{j}_k\pmb{j}_k^T  \times diag\left(\frac{1}{\sigma^2 + \lambda^{-1}_{R(i,j)}}\right)}{1+ \sigma^2_b \pmb{j}_k^T  diag\left(\frac{1}{\sigma^2 + \lambda^{-1}_{R(i,j)}}\right) \pmb{j}_k}\]

If we let 
\[\pmb{\ell}_i = \left[ \frac{\sigma^2_b}{\sigma^2 + \lambda^{-1}_{R(i,1)}}, \frac{\sigma^2_b}{\sigma^2 + \lambda^{-1}_{R(i,2)}}, \cdots \frac{\sigma^2_b}{\sigma^2 + \lambda^{-1}_{R(i,k)}}  \right]\]
then 
\[(W^{-1}+ZGZ^T)_{i}^{-1} = diag\left(\frac{1}{\sigma^2 + \lambda^{-1}_{R(i,j)}}\right) + \frac{\pmb{\ell}_i \pmb{\ell}_i^T}{\sigma^2_b (1+(\pmb{\ell}^{1/2}_i)^T \pmb{\ell}_i^{1/2})}\]

Next, we can add the additional components that are not dependent on $X$. So
\begin{align*}
W - WZ(ZWZ^T+G^{-1})^{-1}Z^TW &= W - WZ(G-GZ^T(W^{-1}+ZGZ^T)^{-1}ZG)Z^TW\\
&= W - WZGZ^TW + WZGZ^T  (W^{-1}+ZGZ^T)^{-1} (WZGZ^T)^T
\end{align*}
Since $ZGZ^T = [\sigma^2_b I_b \otimes \pmb{j}_i \pmb{j}_i^T +\sigma^2 I_{bk}]$, we have
\begin{align*}
WZGZ^TW &=\left[ \begin{array}{cccc}
\sigma^2_b \pmb{\lambda}_1 \pmb{\lambda}_1^T & \pmb{0}                                                           & \cdots & \pmb{0} \\
\pmb{0}                                                           & \sigma^2_b \pmb{\lambda}_2 \pmb{\lambda}_2^T & \cdots & \pmb{0} \\
\vdots                                                              &\vdots                                                               & \ddots & \vdots\\
\pmb{0}                                                           & \pmb{0}                                                           & \cdots & \sigma^2_b \pmb{\lambda}_b \pmb{\lambda}_b^T 
\end{array} \right] \times \sigma^2 W^2\\
WZGZ^T    &= \left[ \begin{array}{cccc}
\sigma^2_b \pmb{\lambda}_1 \pmb{j}_k^T & \pmb{0}                                                    & \cdots & \pmb{0} \\
\pmb{0}                                                 & \sigma^2_b \pmb{\lambda}_2 \pmb{j}_k^T & \cdots & \pmb{0} \\
\vdots                                                    &\vdots                                                        & \ddots & \vdots\\
\pmb{0}                                                 & \pmb{0}                                                    & \cdots & \sigma^2_b \pmb{\lambda}_b \pmb{j}_k^T 
\end{array} \right] \times \sigma^2 W\\
\end{align*}
where $\pmb{\lambda}_i = (\lambda_{R(i,1)},\lambda_{R(i,2)},\cdots,\lambda_{R(i,k)})^T$. Since each of these matrices are block diagonal, the $ (i,i)^{\textrm{th}} $ block of $W - WZ(ZWZ^T+G^{-1})^{-1}Z^TW$ becomes
\begin{align*}
\hspace{2em}&\hspace{-2em} (W - WZ(ZWZ^T+G^{-1})^{-1}Z^TW)_{i} \\
&= W-WZGZ^TW + WZGZ^T \times (W^{-1}+ZGZ^T)^{-1} \times (WZGZ^T)^T\\
&= diag(\lambda_{R(i,j)}) - (\sigma^2_b \pmb{\lambda}_i \pmb{\lambda}_i^T + \sigma^2 diag(\lambda_{R(i,j)})^2)\\ 
& \hspace{1cm}    + (\sigma^2_b \pmb{\lambda}_i \pmb{j}_k^T + \sigma^2 diag(\lambda_{R(i,j)}))     
\times \left(diag\left(\frac{1}{\sigma^2 + \lambda^{-1}_{R(i,j)}}\right) 
- \frac{\pmb{\ell}_i \pmb{\ell}_i^T}{\sigma^2_b (1+(\pmb{\ell}_i^{1/2})^T\pmb{\ell}_i^{1/2})}\right)   
\times (\sigma^2_b \pmb{\lambda}_i \pmb{j}_k^T + \sigma^2 diag(\lambda_{R(i,j)}))^T\\
&= diag(\lambda_{R(i,j)} - \sigma^2 \lambda_{R(i,j)}^2 ) - \sigma^2_b \pmb{\lambda}_i \pmb{\lambda}_i^T\\
& \hspace{1cm} + \left(\sigma^2_b \pmb{\lambda}_i \pmb{j}_k^T \times diag\left(\frac{1}{\sigma^2 + \lambda^{-1}_{R(i,j)}}\right)
- \sigma^2_b \pmb{\lambda}_i \pmb{j}_k^T \times \frac{\pmb{\ell}_i \pmb{\ell}_i^T}{\sigma^2_b (1+(\pmb{\ell}_i^{1/2})^T\pmb{\ell}_i^{1/2})}
+ \sigma^2 diag(\lambda_{R(i,j)}) \times diag\left(\frac{1}{\sigma^2 + \lambda^{-1}_{R(i,j)}}\right)\right.\\
& \hspace{2cm} \left. - \sigma^2 diag(\lambda_{R(i,j)}) \times \frac{\pmb{\ell}_i \pmb{\ell}_i^T}{\sigma^2_b (1+(\pmb{\ell}_i^{1/2})^T\pmb{\ell}_i^{1/2})}\right) \times (\sigma^2_b \pmb{\lambda}_i \pmb{j}_k^T + \sigma^2 diag(\lambda_{R(i,j)}))^T\\
&= diag(\lambda_{R(i,j)} - \sigma^2 \lambda_{R(i,j)}^2 ) - \sigma^2_b \pmb{\lambda}_i \pmb{\lambda}_i^T\\
& \hspace{1cm} + \left(\pmb{\lambda}_i \pmb{\ell}_i^T
- \frac{\pmb{\lambda}_i \pmb{j}_k^T \pmb{\ell}_i \pmb{\ell}_i^T}{(1+(\pmb{\ell}_i^{1/2})^T\pmb{\ell}_i^{1/2})}
+ diag\left(\frac{\sigma^2\lambda_{R(i,j)}}{\sigma^2 + \lambda^{-1}_{R(i,j)}}\right) - 
\frac{ \sigma^2\pmb{m}_i \pmb{\ell}_i^T}{\sigma^2_b (1+(\pmb{\ell}_i^{1/2})^T\pmb{\ell}_i^{1/2})}\right)  
\times (\sigma^2_b \pmb{\lambda}_i \pmb{j}_k^T + \sigma^2 diag(\lambda_{R(i,j)}))^T\\ 
&= diag(\lambda_{R(i,j)} - \sigma^2 \lambda_{R(i,j)}^2 ) - \sigma^2_b \pmb{\lambda}_i \pmb{\lambda}_i^T
+ \sigma^2_b \pmb{\lambda}_i \pmb{\ell}_i^T \pmb{j}_k \pmb{\lambda}_i^T
- \sigma^2_b \frac{\pmb{\lambda}_i \pmb{j}_k^T \pmb{\ell}_i \pmb{\ell}_i^T \pmb{j}_k \pmb{\lambda}_i^T}{(1+(\pmb{\ell}_i^{1/2})^T\pmb{\ell}_i^{1/2})} + \sigma^2 \pmb{m}_i \pmb{\lambda}_i^T\\
& \hspace{1cm}   - \frac{ \sigma^2\pmb{m}_i \pmb{\ell}_i^T \pmb{j}_k \pmb{\lambda}_i^T}{(1+(\pmb{\ell}_i^{1/2})^T\pmb{\ell}_i^{1/2})} 
+ \sigma^2 \pmb{\lambda}_i \pmb{m}_i^T
- \frac{\sigma^2 \pmb{\lambda}_i \pmb{j}_k^T \pmb{\ell}_i \pmb{m}_i^T}{(1+(\pmb{\ell}_i^{1/2})^T\pmb{\ell}_i^{1/2})} + diag\left(\frac{\sigma^4\lambda^2_{R(i,j)}}{\sigma^2 + \lambda^{-1}_{R(i,j)}}\right)
- \frac{ \sigma^4\pmb{m}_i \pmb{m}_i^T}{\sigma^2_b (1+(\pmb{\ell}_i^{1/2})^T\pmb{\ell}_i^{1/2})}
\end{align*}
where 
\[\pmb{m}_i = \left( \frac{\sigma^2_b \lambda_{R(i,1)}}{\sigma^2 + \lambda^{-1}_{R(i,1)}}, \frac{\sigma^2_b \lambda_{R(i,2)}}{\sigma^2 + \lambda^{-1}_{R(i,2)}}, \cdots \frac{\sigma^2_b \lambda_{R(i,k)}}{\sigma^2 + \lambda^{-1}_{R(i,k)}}  \right)^T\]
Since $ \pmb{\ell}_i^T \pmb{j}_k = (\pmb{\ell}_i^{1/2})^T \pmb{\ell}_i^{1/2}$ and $ \pmb{j}_k^T \pmb{\ell}_i = (\pmb{\ell}_i^{1/2})^T \pmb{\ell}_i^{1/2} $, which are constants, we obtain

\begin{align*}
\hspace{2em}&\hspace{-2em}(W - WZ(ZWZ^T+G^{-1})^{-1}Z^TW)_i\\
&= diag(\lambda_{R(i,j)} - \sigma^2 \lambda_{R(i,j)}^2 ) - \sigma^2_b \pmb{\lambda}_i \pmb{\lambda}_i^T
+ \sigma^2_b \pmb{\lambda}_i (\pmb{\ell}_i^{1/2})^T \pmb{\ell}_i^{1/2} \pmb{\lambda}_i^T
- \sigma^2_b \frac{\pmb{\lambda}_i (\pmb{\ell}_i^{1/2})^T \pmb{\ell}_i^{1/2} (\pmb{\ell}_i^{1/2})^T \pmb{\ell}_i^{1/2} \pmb{\lambda}_i^T}{(1+(\pmb{\ell}_i^{1/2})^T\pmb{\ell}_i^{1/2})} + \sigma^2 \pmb{m}_i \pmb{\lambda}_i^T\\
& \hspace{1cm}   - \frac{ \sigma^2\pmb{m}_i (\pmb{\ell}_i^{1/2})^T \pmb{\ell}_i^{1/2} \pmb{\lambda}_i^T}{(1+(\pmb{\ell}_i^{1/2})^T\pmb{\ell}_i^{1/2})} 
+ \sigma^2 \pmb{\lambda}_i \pmb{m}_i^T
- \frac{\sigma^2 \pmb{\lambda}_i (\pmb{\ell}_i^{1/2})^T \pmb{\ell}_i^{1/2} \pmb{m}_i^T}{(1+(\pmb{\ell}_i^{1/2})^T\pmb{\ell}_i^{1/2})}
+ diag\left(\frac{\sigma^4\lambda^2_{R(i,j)}}{\sigma^2 + \lambda^{-1}_{R(i,j)}}\right)
- \frac{ \sigma^4\pmb{m}_i \pmb{m}_i^T}{\sigma^2_b (1+(\pmb{\ell}_i^{1/2})^T\pmb{\ell}_i^{1/2})}\\
&= diag\left(\lambda_{R(i,j)} - \sigma^2 \lambda^2_{R(i,j)} + \frac{\sigma^4\lambda^2_{R(i,j)}}{\sigma^2 + \lambda^{-1}_{R(i,j)}} \right) 
+\left(\sigma^2_b (\pmb{\ell}_i^{1/2})^T \pmb{\ell}_i^{1/2}  - \sigma^2_b - \frac{\sigma^2_b  \pmb{\ell}_i^T \pmb{\ell}_i}{(1+(\pmb{\ell}_i^{1/2})^T\pmb{\ell}_i^{1/2})}  \right) \pmb{\lambda}_i \pmb{\lambda}_i^T\\
& \hspace{1cm} + \left(\sigma^2  - \frac{ \sigma^2 (\pmb{\ell}_i^{1/2})^T \pmb{\ell}_i^{1/2} }{(1+(\pmb{\ell}_i^{1/2})^T\pmb{\ell}_i^{1/2})}   \right) \pmb{m}_i \pmb{\lambda}_i^T
+ \left(\sigma^2 - \frac{\sigma^2(\pmb{\ell}_i^{1/2})^T \pmb{\ell}_i^{1/2}}{(1+(\pmb{\ell}_i^{1/2})^T\pmb{\ell}_i^{1/2})} \right) \pmb{\lambda}_i \pmb{m}_i^T - \frac{ \sigma^4\pmb{m}_i \pmb{m}_i^T}{\sigma^2_b (1+(\pmb{\ell}_i^{1/2})^T\pmb{\ell}_i^{1/2})}\\
&= diag\left(\frac{1}{\sigma^2 + \lambda^{-1}_{R(i,j)}} \right) 
-\left(\frac{\sigma^2_b}{(1+(\pmb{\ell}_i^{1/2})^T\pmb{\ell}_i^{1/2})}  \right) \pmb{\lambda}_i \pmb{\lambda}_i^T
+ \left(\frac{\sigma^2}{(1+(\pmb{\ell}_i^{1/2})^T\pmb{\ell}_i^{1/2})}   \right) \pmb{m}_i \pmb{\lambda}_i^T
+ \left(\frac{\sigma^2}{(1+(\pmb{\ell}_i^{1/2})^T\pmb{\ell}_i^{1/2})} \right) \pmb{\lambda}_i \pmb{m}_i^T\\
& \hspace{1cm} - \frac{ \sigma^4\pmb{m}_i \pmb{m}_i^T}{\sigma^2_b (1+(\pmb{\ell}_i^{1/2})^T\pmb{\ell}_i^{1/2})}\\
&= diag\left(\frac{1}{\sigma^2 + \lambda^{-1}_{R(i,j)}} \right) 
+ \frac{1}{\sigma^2_b(1+(\pmb{\ell}_i^{1/2})^T\pmb{\ell}_i^{1/2})}
\left(-\sigma^4_b \pmb{\lambda}_i \pmb{\lambda}_i^T + \sigma^2 \sigma^2_b \pmb{m}_i \pmb{\lambda}_i^T
+  \sigma^2 \sigma^2_b \pmb{\lambda}_i \pmb{m}_i^T -\sigma^4\pmb{m}_i \pmb{m}_i^T  \right)\\
&= diag\left(\frac{1}{\sigma^2 + \lambda^{-1}_{R(i,j)}} \right) 
- \frac{1}{\sigma^2_b(1+(\pmb{\ell}_i^{1/2})^T\pmb{\ell}_i^{1/2})}
\left(\sigma^2 \pmb{m}_i - \sigma^2_b \pmb{\lambda} \right) \left(\sigma^2 \pmb{m}_i - \sigma^2_b \pmb{\lambda} \right)^T
\end{align*}

Since
\begin{align*}
\sigma^2 \pmb{m}_i - \sigma^2_b \pmb{\lambda} &= \left( \frac{\sigma^2\sigma^2_b \lambda_{R(i,1)}}{\sigma^2 + \lambda^{-1}_{R(i,1)}}, \frac{\sigma^2\sigma^2_b \lambda_{R(i,2)}}{\sigma^2 + \lambda^{-1}_{R(i,2)}}, \cdots \frac{\sigma^2 \sigma^2_b \lambda_{R(i,k)}}{\sigma^2 + \lambda^{-1}_{R(i,k)}}  \right)^T - \left( \sigma^2_b\lambda_{R(i,1)},  \sigma^2_b\lambda_{R(i,2)}, \cdots , \sigma^2_b\lambda_{R(i,k)}  \right)^T\\
&=  \left( \frac{\sigma^2\sigma^2_b \lambda_{R(i,1)} - \sigma^2\sigma^2_b \lambda_{R(i,1)} - \sigma^2_b}{\sigma^2 + \lambda^{-1}_{R(i,1)}}, \frac{\sigma^2\sigma^2_b \lambda_{R(i,2)} - \sigma^2\sigma^2_b \lambda_{R(i,2)} - \sigma^2_b}{\sigma^2 + \lambda^{-1}_{R(i,2)}}, \cdots \frac{\sigma^2 \sigma^2_b \lambda_{R(i,k)} - \sigma^2 \sigma^2_b \lambda_{R(i,k)} - \sigma^2_b}{\sigma^2 + \lambda^{-1}_{R(i,k)}}  \right)^T \\
&= - \left( \frac{\sigma^2_b}{\sigma^2 + \lambda^{-1}_{R(i,1)}}, \frac{\sigma^2_b}{\sigma^2 + \lambda^{-1}_{R(i,2)}}, \cdots \frac{\sigma^2_b}{\sigma^2 + \lambda^{-1}_{R(i,k)}}  \right)^T\\
&= -\pmb{\ell}_i,
\end{align*}
we can simplify to obtain
\[(W - WZ(ZWZ^T+G^{-1})^{-1}Z^TW)_i = diag\left(\frac{1}{\sigma^2 +\lambda^{-1}_{R(i,j)}} \right) 
- \frac{\pmb{\ell}_i \pmb{\ell}_i^T}{\sigma^2_b(1+(\pmb{\ell}_i^{1/2})^T\pmb{\ell}_i^{1/2})}\]
It follows that $M^{\textrm{marg}}_{\pmb \beta}(\xi,\pmb{\beta},\sigma,\sigma_b)$ is block diagonal with the $(i,i)$ block, $M^{\textrm{marg}}_{\pmb \beta}(\xi,\pmb{\beta},\sigma,\sigma_b)_i$, given by
\[M^{\textrm{marg}}_{\pmb \beta}(\xi,\pmb{\beta},\sigma,\sigma_b)_i = X_i^T \left( \diag\left(\frac{1}{\sigma^2 +\lambda^{-1}_{R(i,j)}} \right) 
- \frac{\pmb{\ell}_i \pmb{\ell}_i^T}{\sigma^2_b(1+(\pmb{\ell}_i^{1/2})^T\pmb{\ell}_i^{1/2})} \right) X_i,\]
where $X_i$ contains the rows if the design matrix corresponding to block $i$.

%  To get the journal style of heading for an appendix, mimic the following.

%\section{}
%\subsection{Title of appendix}
%
%Put your short appendix here.  Remember, longer appendices are
%possible when presented as Supplementary Web Material.  Please 
%review and follow the journal policy for this material, available
%under Instructions for Authors at \texttt{http://www.biometrics.tibs.org}.

\end{document}